# Room temperature single-photon terahertz detection with thermal Rydberg atoms

Danyang Li[1,†], Zhengyang Bai[1,†], Xiaoliang Zuo[1], Yuelong Wu[1], Jiteng Sheng[1,2,*], and Haibin Wu[1,2,3,*]

[1]State Key Laboratory of Precision Spectroscopy, Institute of Quantum Science and Precision Measurement, East China Normal University, Shanghai 200062, China

[2]Collaborative Innovation Center of Extreme Optics, Shanxi University, Taiyuan 030006, China

[3]Shanghai Research Center for Quantum Sciences, Shanghai 201315, China

[†]These authors contributed equally to this work.

*jtsheng@lps.ecnu.edu.cn; hbwu@phy.ecnu.edu.cn

**Single-photon terahertz (THz) detection is one of the most demanding technology for a variety of fields and could lead to many breakthroughs. Although its significant progress has been made in the last two decades, operating it at room temperature still remains a great challenge. Here, we demonstrate, for the first time, the room temperature THz detector at single-photon levels based on nonlinear wave mixing in thermal Rydberg atomic vapor. The low-energy THz photons are coherently upconverted to the high-energy optical photons via a nondegenerate Rydberg state involved six-wave-mixing process, and therefore, the single-photon THz detection is achieved by a conventional optical single-photon counting module. The noise equivalent power of such a detector is reached to be $9.5 \times 10^{-19}$ W/Hz$^{1/2}$, which is more than four orders of magnitude lower than the state-of-the-art room temperature THz detectors. The optimum quantum efficiency of the whole wave-mixing process is about 4.3% with 40.6 dB dynamic range, and the maximum conversion bandwidth is 172 MHz, which is all-optically controllable. The developed fast and continuous-wave single-photon THz detector at room temperature operation has a great potential to be portable and chip-scale, and could be revolutionary for a wide range of applications in remote sensing, wireless communication, biomedical diagnostics, and quantum optics.**

Terahertz (THz) in a frequency range of $10^{12\pm1}$ Hz has been considered as one of the top ten technologies that will change the world of the future [1-3]. Owing to many unique properties over the electromagnetic waves in other frequency bands, it plays an increasingly important role in fields as diverse as astronomy, security, communication, and biomedicine. Among these applications, ultrahigh sensitive THz detection to the single-photon level sensitivity is crucial for the rich far-infrared spectroscopic research, THz radar, and beyond-5G wireless communications [4,5]. However, a "THz gap" [6,7], for which the technology in the frequency domain is relatively absent compared to the optical and microwave regions, leads to realize the THz single-photon detectors very demanding. Although some quantum detectors, such as quantum dot detectors [8], quantum wells detectors [9], and quantum capacitance detectors [10], have been reported to have sensitivity to achieve THz single-photon detection, all of these techniques developed thus far have to be required a cryogenic environment.

Room temperature operation is critically important for real-world applications, and alternative protocols have been developed, such as plasmonic photomixing [11], field-effect transistors [12], topological insulators [13], photothermoelectric effect in graphene [14], electromagnetic induced well [15], etc. The upconversion of THz radiation to visible/near-infrared is one of the most promising highly sensitive room temperature detection approaches in these THz detectors. Such a strategy has been utilized to detect mid-infrared photons with nonlinear crystals [16-18] and optomechanics [19,20]. In the THz region, various nonlinear media have been used, such as quantum dots and bulk nonlinear crystals [21-25], however, the detection sensitivity is currently many orders of magnitude away from the single-photon level. Therefore, to find a suitable room temperature nonlinear medium with high quantum efficiency and low noise is an active goal in the field.

Here we use thermal Rydberg atoms for the first time to realize the room temperature single-photon THz detection. Rydberg atoms in highly excited states have many extraordinary characteristics, including exaggerated sensitivity to external fields, large dipole-dipole interactions, and giant nonlinearities [26]. Many previous efforts have been mainly devoted to investigate Rydberg atoms as perfect quantum systems for precision measurements of microwave electric fields [27-30], quantum many-body simulators [31,32], and deterministic single photon sources [33,34]. Recently, coherent microwave-to-optics conversion in cold Rydberg atoms have been demonstrated with a high conversion efficiency [35-37] due to the fact that cold atomic ensembles exhibit large cooperativity and nonlinearity for the phase matching condition. However, the relatively complicated setup and non-continuous measurements with large dwell time are insurmountable difficulties for laser-cooled atoms.

We overcome these challenges to achieve a highly effective THz-to-optical frequency converter by using a thermal rubidium atomic vapor as the nonlinear medium. We demonstrate that the large atomic coherence induced by the electromagnetic fields interacting with the Rydberg states incorporating both THz and optical transitions enables an optimum quantum efficiency 4.3% even under large Doppler broadening. Our upconversion scheme operates with continuous-wave (cw) weak THz signals at room temperature, leading single-photon THz detection by measuring the upconverted visible photons with a silicon single photon detector without requiring deep cryogenic cooling. The noise equivalent power (NEP) of such a detector is $9.5\times 10^{-19}$ W/Hz$^{1/2}$, which is over four orders of magnitude smaller than other reported room temperature THz detectors. The conversion bandwidth can be as broad as 172 MHz with a relatively strong THz signal, which is significantly larger than the bandwidth for microwave-to-optics conversion in cold atoms [35-37]. Moreover, we use such a fast and coherent conversion detection to measure the statistical distribution of the input THz photons, which is not possible for the Rydberg electromagnetically-induced-transparency (EIT) based detection schemes [27-29], and demonstrate its capability to perform the single photon detection.

The illustration of the room temperature single-photon THz detection is shown in Fig. 1a, where the key component is a THz-to-optics converter. The input low-energy THz photons with a frequency $\nu_T \sim 0.107$ THz are upconverted to the high-energy signal photons ($\nu_S \sim 384.5$ THz) through a six-wave mixing (6WM) process with three auxiliary optical fields $A_i$ (i=1,2,4) and a microwave field ($A_3$) in a thermal Rydberg $^{87}$Rb gas. The corresponding field-coupled six-level atomic system is exhibited in Fig. 1b. $|1\rangle$ is the ground state, $|2\rangle$ and $|6\rangle$ are the intermediate states, and $|3\rangle$, $|4\rangle$ and $|5\rangle$ are the Rydberg states. The frequency and wavevector of the signal field are determined by the energy conservation and phase matching condition, which are $\nu_S = \nu_T + \nu_1 + \nu_2 + \nu_3 - \nu_4$ and $\boldsymbol{k}_S = \boldsymbol{k}_T + \boldsymbol{k}_1 + \boldsymbol{k}_2 + \boldsymbol{k}_3 - \boldsymbol{k}_4$, respectively. Here $\nu_i$ and $\boldsymbol{k}_i$ (i=S,T,1,2,3,4) are the frequency and wavevector of the corresponding field. The combination of multilevel atomic coherence and large Rydberg dipole moment [26] enable the high efficiency 6WM process in free space at room temperature.

A schematic of the experimental setup is displayed in Fig. 1c. $A_1$, $A_2$ and $A_4$ are the cw optical fields derived from three different lasers with wavelengths ~ 795 nm, 476 nm, and 482 nm, respectively. They propagate coaxially and focus onto the center of a 5 mm long rubidium atomic vapor cell, which is heated up to ~ 120 °C. $A_3$ is a microwave field with a frequency of 62.3 GHz. $A_3$ and THz fields are generated by the horn antennas and co-propagate with the optical auxiliary fields at a small angle. The signal field generated continuously through the nondegenerate 6WM process has a wavelength of ~ 780 nm, which is spatially separated and filtered by a blazed grating and a single mode fiber at a high extinction ratio, and subsequently measured with a single photon counting module (SPCM). The transmission of $A_1$ ($T_1$) is monitored by a photomultiplier tube (PMT). The generated signal strength is maximized by scanning the frequency of $A_1$ and optimizing the parameters of the auxiliary fields, such as frequency detunings and Rabi frequencies. We find that the best quantum efficiency is realized in a slightly off-resonant of $A_1$ field around $\Delta_1 = -2\pi \times 5.2$ MHz mainly due to the Rabi splitting at the modest auxiliary field intensities. The typical spectra of $A_1$ and signal fields is presented in Fig. 1d. The resonance of transmission $T_1$ (the blue curve) behaves an EIT-like peak with a Doppler-broadened background. Under phase matching condition, the upconverted signal is generated and has a gain peak (the red curve), which represents the signature of 6WM process.

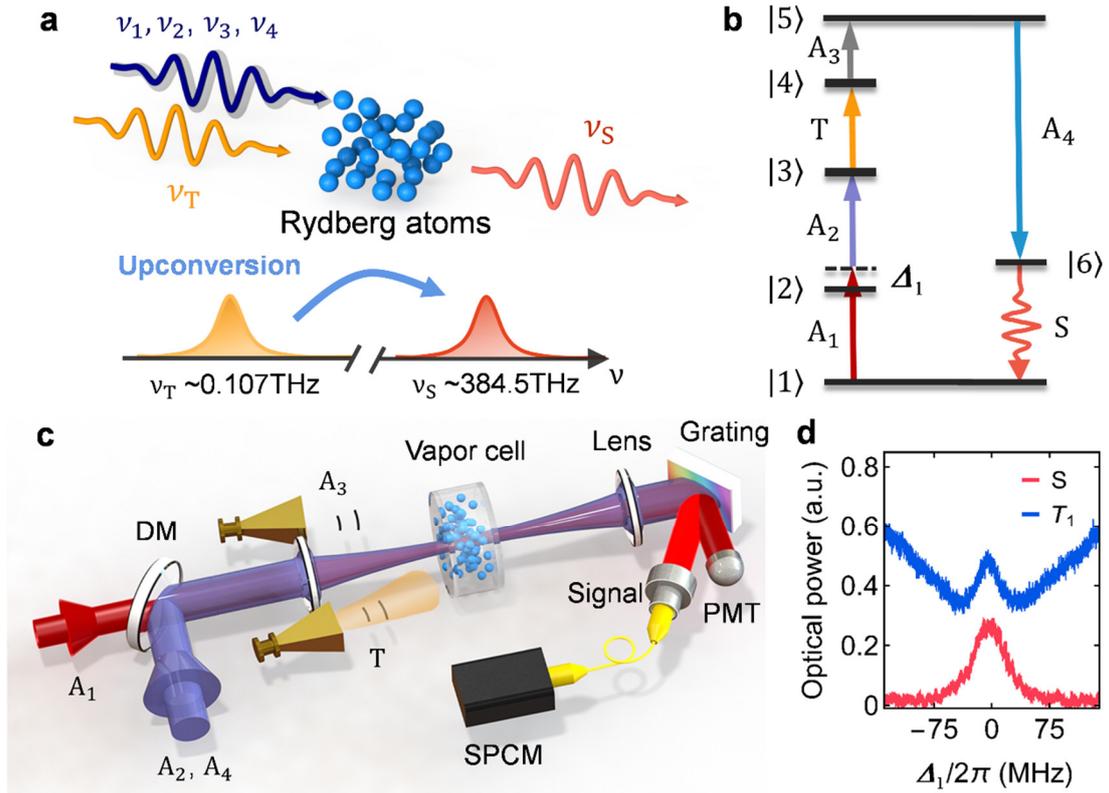

**Fig. 1. Basic principle and experimental setup of single-photon THz detector. a**, Input THz photons are upconverted to the visible optical photons through a nondegenerate 6WM process with the help of four auxiliary fields in thermal Rydberg atoms. **b**, Energy level diagram of a six-level $^{87}$Rb atomic system. $A_1$, $A_2$, $A_3$ and $A_4$ are four auxiliary fields with Rabi frequency $\Omega_i$ (i=1-4) and T represents the THz field. $\Delta_1$ is the frequency detuning of $A_1$ from the atomic transition $|1\rangle \leftrightarrow |2\rangle$. **c**, Experimental setup. DM, dichroic mirror; PMT, photomultiplier tube; SPCM, single photon counting module. **d**, Spectra of the generated 6WM signal (red) and the transmission of $A_1$ ($T_1$) (blue).

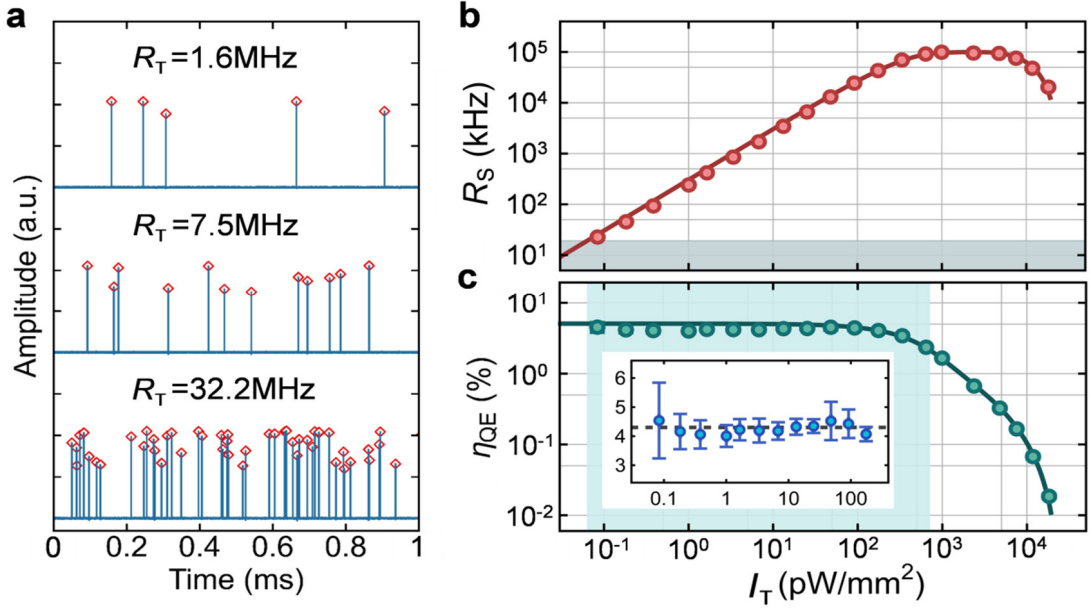

**Fig. 2. Response of THz-to-optics converter. a**, Direct measurements of the upconverted optical photons (include dark count) by the SPCM at three different input THz count rates, i.e., $R_T$ = 1.6, 7.5 and 32.2 MHz, respectively. **b**, Signal count rate of the THz-to-optics converter as a function of the input THz intensity. **c**, Quantum efficiency of the converter as a function of the input THz intensity. The inset shows the linear regime. The solid curves in **b** and **c** are numerical simulations.

The performance of THz-to-optics converter is experimentally characterized. Figure 2a presents the direct measurements of the upconverted optical photons by the SPCM at three different input THz field count rates (intensities). When the number of input THz photons is relatively low, the upconverted optical photon rate grows linearly by increasing the THz intensity, which is referred as the linear response regime (i.e., $R_S \propto I_T$ ). Here, $R_S$ is the count rate of the signal field, and $I_T$ represents the intensity of THz field. The input THz field is adopted as in the multi-photon coherent state, and THz (upconverted signal) photons can be treated as a classical field with Rabi frequency $\Omega_{T(S)}$. The Rabi frequency of THz field is calibrated by in situ measurement with the Rydberg EIT method [27]. For further increasing power of THz field, the upconverted photon rate gradually reaches saturation and falls off at a high input intensity. In the linear regime, the systemic responding is simple, and it is vital for the THz detection. Outside the linear regime, the THz fields induced strong nonlinearity might lead to rich physics (e.g., THz-driven nonlinear spin control [38] or non-equilibrium phase transition [39]), which is a topic deserving exploration but beyond the scope of the single-photon detector.

The quantum efficiency of the converter is $\eta_{QE} = R_S / R_T$, where $R_T = I_T S_{eff} / h\nu_T$ is the count rate of the input THz field. $S_{eff}$ is the effective area of conversion medium. $R_S$ is inferred by the measurements of the photon count rate at SPCM and the additional losses, i.e. coupling losses and detection efficiency of SPCM ($\eta_{loss} \sim 0.11$). Therefore, the total detection efficiency is $\eta = \eta_{QE}\eta_{loss}$. The maximum quantum efficiency of the converter is realized to be 4.3(0.4) % through meticulous optical alignment and iterative parameter adjustments. It keeps a constant in the linear regime and deceases at relative high input THz intensities (see Fig. 2c). The dynamic range is 40.6 dB, which is defined as the signal-to-noise ratio >1 and -3 dB of the conversion efficiency [40] (the green shaded area in Fig. 2c).

The minimum detectable input THz field is limited by the dark count rate $D$ of the whole detecting system ($D \approx 2$ kHz), which is mainly resulted from the stray optical photons coupled into the SPCM. Therefore, we obtain the NEP of the single photon THz detector at $\text{NEP} = h\nu_T\sqrt{2D}/\eta \approx 9.5 \times 10^{-19} W/Hz^{1/2}$, which is more than four orders of magnitude better than the state-of-the-art highly sensitive room temperature THz detectors [15].

A lower $D$ and a higher $\eta$ are benefit of a lower NEP. $D$ will ultimately be limited by the blackbody radiation at room temperature. $\eta$ can be further enhanced by reducing $\eta_{loss}$ and increasing $\eta_{QE}$. To estimate the highest possible $\eta_{QE}$, we analyze the 6WM dynamics under the weak-excitation condition by adopting the Holstein-Primakoff transformation [41] and obtain the analytical expression

$$\eta_{QE} = G_S^2 G_T^2 \left(-1+e^{-\bar{\alpha}L}\right)^2 / (G_S^2 + G_T^2)^2, \qquad (1)$$

where $G_{S(T)} = g_{S(T)}\Omega_{2(1)}\Omega_{3(4)}$ with the coupling constant $g_{S(T)}$ and $L$ is the length of the conversion medium (see details in Sec. IB of SM). According to Eq. (1), the maximum $\eta_{QE}$ is not restricted to the thermal environment for $\bar{\alpha}L \to \infty$.

$\bar{\alpha} = 2g_S^2(G_S^2 + G_T^2)/(c\Gamma_{th}G_S^2)$ denotes the build-up rate of 6WM, which is inversely proportional to the effective Doppler linewidth $\Gamma_{th}$ [42]. $\Gamma_{th}$ has an energy scale of gigahertz at room temperature. Compared with ultracold atomic systems (with $\Gamma_{th}$ ~ megahertz), a larger $L$ is required to set up 6WM in thermal vapor. However, as $L$ increases, the attenuation of signal field is accumulated and could suppresses 6WM due to the thermal collision and finite Rydberg lifetime (not incorporated in our analytical model). Hence, an effective way to promote $\eta_{QE}$ is to enhance the power of auxiliary field [33, 43]. This is further verified via the exact numerical simulation with a $\eta_{QE}$ up to 91.3% (see Supplementary Materials). Therefore, with reduced $D$ and enhanced $\eta_{QE}$, such a single-photon THz detection with $NEP \sim 10^{-21} W/Hz^{1/2}$ could be expected, which is comparable to the state-of-the-art single-photon THz detectors operating at cryogenic temperatures.

The conversion bandwidth is another essential index for the performance of a detector, and it is typically inversely proportional to the detector's speed. Figure 3a shows the conversion bandwidth of the THz detector at three different powers of field $A_1$. At a relatively small power, the signal spectrum exhibits a single Lorentz peak. By increasing the power of $A_1$, the signal peak splits into two peaks due to the Rabi splitting, and the splitting is monotonically depending on the power of $A_1$. Besides, due to coherent population trapping (with $\Omega_1 > \Omega_2$), a large fraction of atoms is transferred into Rydberg states. This leads to significant Rydberg collisions [44, 45]. Both Rabi splitting and thermal broadening can contribute to the enhancement of bandwidth for thermal gas. The conversion bandwidth is defined as the full-width at half-maximum (FWHM) of the signal strength as shown in Fig. 3a. As one can see in Fig. 3b, the bandwidth of the THz detector can be all-optically tuned by simply changing the power of the auxiliary light field. This is especially useful for the situations where the bandwidth of the detector needs to be controllable. Based on our setting, the maximum bandwidth has been achieved in 2π×172 MHz. This value is at least more than one order of magnitude larger than the cold atoms based schemes [35], and is sufficient for real-time video signal transfer [46].

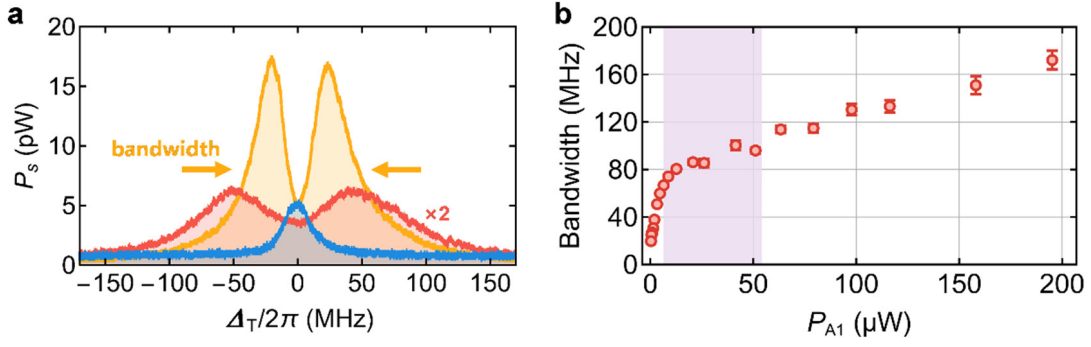

**FIG. 3. Conversion bandwidth of the THz detector. a,** Generated signal spectra at three different power of $A_1$. The blue, orange, and red curves correspond to the power ~ 0.5, 12.7, and 195.2 μW, respectively. **b,** Bandwidth as a function of the power of $A_1$. The error bars are the standard deviations of five independent measurements. The purple shaded region indicates that the central dip of the signal spectra is less than the half of its maximum.

One of the most important features of the fast and high efficiency single photon detection technique is the ability to perform the experiments in quantum optics with unprecedented measurement precision. For instance, coincidence measurements can be used to investigate the nonclassical property of light. In order to demonstrate that the THz-to-optics converter can maintain the statistics of input THz photons, we measure the second-order degree of coherence $g_S^{(2)}$ based on the Hanbury-Brown-Twiss (HBT) technique.

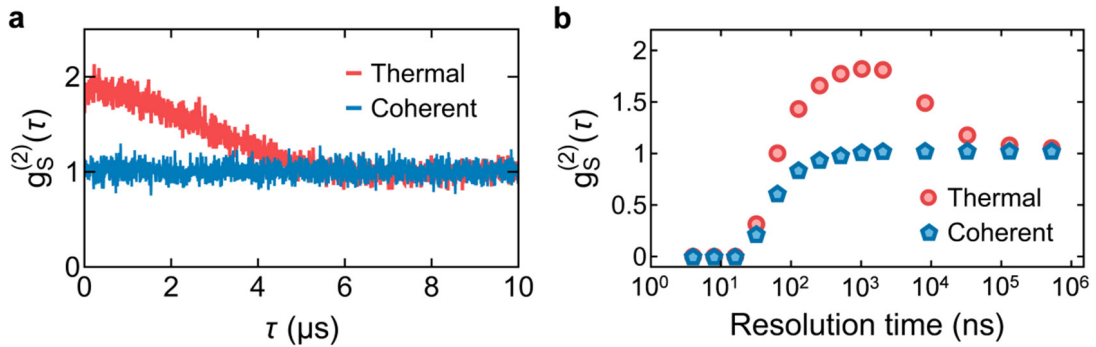

**FIG. 4. Second-order degree of coherence measurements. a,** Measured $g_S^{(2)}(\tau)$ of the upconverted signal photons for the coherent and thermal input THz fields via HBT scheme. **b,** Measured $g_S^{(2)}(0)$ of the upconverted signal photons as a function of resolution time via a single visible-light detector.

Figure 4a shows the measured $g_S^{(2)}(\tau)$ of the upconverted signal photons for both coherent and thermal input THz fields. For the coherent input of THz field, the signal field is generated with $g_S^{(2)}(0) = 1$. For the pseudothermal input THz field, a thermal signal field with $g_S^{(2)}(0) \approx 2$ is obtained. This indicates that the quantum statistics of THz can be maintained into optical photons. This is a unique property for our system, where exotic statistics characteristics can be probed via 6WM process, which is not possible for the Rydberg EIT based detection schemes. As a result, one might anticipate that such a THz detector could be particular helpful when quantum THz sources become available in the future. By using simple autocorrelation method, the properties of a single photon detector can be extracted [47]. The zero-delay second-order degree of coherence $g_S^{(2)}(0)$ is measured in Fig. 4b. When the resolution time is smaller than 32 ns, $g_S^{(2)}(0) = 0$ for both coherent and thermal input THz fields. By bypassing the THz-to-optics converter, we find that the recovery time of the THz single photon detector is the same as the commercial silicon single photon detector, which indicates that the dynamic response is currently limited by the commercial visible-light single-photon detectors rather than the Rydberg atoms based nonlinear wave mixing process.

In conclusion, we have developed a room temperature single photon THz detector based on the THz-to-optics upconversion via a nondegenerate Rydberg state involved nonlinear wave mixing process. The superior performance in the NEP and continuous-wave detection speed demonstrates that Rydberg atomic THz convertor offers opportunities for measuring THz fields down to single photon levels. Combined with the potential for portability and extensive wavelength coverage, such a THz detection based on thermal Rydberg atoms not only promises a breakthrough in future technology but also open an avenue for a multitude of transformative applications, ranging from communication and imaging to spectroscopy and beyond.